# Analysis of Enhancement factors of Parity and Time Reversal Violating Effects for Monofluorides

A. Sunaga*,[1] V. S. Prasannaa,[2] M. Abe,[1] M. Hada,[1] and B. P. Das[3]

[1]*Tokyo Metropolitan University, 1-1, Minami-Osawa, Hachioji-city, Tokyo 192-0397, Japan*

[2]*Indian Institute of Astrophysics, II Block, Koramangala, Bangalore 560 034, India*
*Department of Physics, Calicut University, Malappuram, Kerala 673 635, India*

[3]*Department of Physics and International Education and Research Center of Science*
*Tokyo Institute of Technology, 2-12-1-H86 Ookayama, Meguro-ku, Tokyo152-8550, Japan*

Heavy polar diatomic molecules are currently one of the leading candidates for probing physics beyond the Standard Model via studies of time-reversal ($T$) and parity ($P$) violations. In this work, we analyze the effective electric field ($E_{\text{eff}}$) that is required for determining the electron electric dipole moment (eEDM), and the scalar-pseudoscalar (S-PS) interaction constant ($W_s$), in group 12 and group 2 systems. We use a relativistic coupled cluster method for our calculations, and find that group 12 monofluorides have large $E_{\text{eff}}$ and $W_s$ (for example, the values of $E_{\text{eff}}$ and $W_s$ of CnF, the heaviest group 12 fluoride, are 662 GV/cm and 3360 kHz, respectively). The reason for this is the contraction of the valence $s$ and $p$ orbitals due to the weak screening effect of the outermost core's $d$ electron. The calculations of $E_{\text{eff}}$ and $W_s$ show that their ratio, ($W_s/E_{\text{eff}}$), increases with Z. Based on these results, as well as experimental suitability, we propose SrF and CdF as new candidate molecules for experiment.

The electric dipole moment of the electron (eEDM, or $d_e$) is a consequence of simultaneous violations of parity ($P$) and time reversal ($T$) symmetries [1-4]. It is of great interest in constraining theories beyond the Standard Model (BSM) of particle physics [5,6], and also in providing insights into the matter-antimatter asymmetry in the Universe [7,8]. To extract the eEDM, experiments use heavy polar diatomic molecules [9-11].

Also of interest in fundamental physics is the $P$ and $T$ violating scalar-pseudoscalar (S-PS) interaction. This is a type of interaction between electrons and nuclei, but it requires not a scalar Higgs like in the Standard Model, but a Higgs with scalar and pseudoscalar components, for example, in the aligned two-Higgs doublet model [12]. This interaction is also $T$ violating like the eEDM, and can shed light on the baryon asymmetry in the Universe. The coupling constant associated with this interaction is the S-PS interaction constant, $k_s$, and determining it is important for BSM physics, just as in the case of $d_e$.

In an experiment on a paramagnetic molecule, one measures the shift in its energy, due to $d_e$ and $k_s$. The energy shift due to eEDM is given by the negative of the product of $d_e$ and an effective electric field, $E_{\text{eff}}$. In the case of the S-PS interaction, the quantity analogous to $E_{\text{eff}}$ is the S-PS coefficient, $W_s$. The energy shifts that are experimentally measured have been smaller than their errors so far, and therefore, the eEDM and S-PS interaction have not been discovered yet. Since an experiment actually measures both the eEDM and the S-PS interaction, one actually sets one quantity to zero, and obtains a bound on the other. However, if we measure two energy shifts using two *different* systems, both the quantities can be obtained without assuming that the other is zero.

Molecules with larger $E_{\text{eff}}$ and $W_s$ are experimentally more favorable because they lead to higher sensitivity. According to semi-empirical formulas for atomic enhancement factor $K$ [13] and molecular $E_{eff}$ [14], not only $Z$ but also the effective radial quantum number $\nu$ contributes to $K$ and $E_{eff}$. $\nu$ is directly related to the net screened charge experienced by the unpaired electron, and this mainly depends on the group of the periodic table. An example of the screening effect is the contraction of the valence $s$ orbital for group 12 atoms, arising from the weak screening by the outermost $d$ electrons [15]. A study on the enhancement factor associated with the anapole moment with $Z$ was performed, on group 12 monohydrides [16]. However, to the best of our knowledge, the contribution of the screening effect to $E_{eff}$ and $W_s$ has never been clearly discussed previously.

In this work, we calculate $E_{eff}$ and $W_s$ of XF molecules (X = Sr, Cd, Ba, Yb, Hg, Ra and Cn) using the Dirac-Fock (DF) and relativistic Coupled Cluster Singles and Doubles (CCSD) methods.

Our results show that the group 12 monofluorides, CdF and HgF, have larger values of $E_{eff}$ and $W_s$ than their heavier group 2 counterparts, BaF and RaF, respectively. We explain the reason of the enhancement for group 12 monofluorides from the viewpoint of the contraction of the valence wavefunction. We confirm the occurrence of this contraction, by performing atomic orbital calculations, using the GRASP2K package [17]. The contraction is due to the weak screening effect of $(n-1)d$ electrons, where $n$ is principal quantum number of the valence orbital. We also show that the departure from the expected $Z$ dependence in group 12 and 2 fluoride molecules is only found in molecular $E_{eff}$ and $W_s$, and not in the atomic enhancement factors of group 12 and 2 cations.

We also find from our results that the ratio $W_s/E_{eff}$ itself monotonically increases with the atomic number $Z$ of the heavier atom, although $E_{eff}$ and $W_s$ do not. Based on the result, we propose two new candidate molecules for experiment, SrF and CdF. We discuss their suitability based on a recent experimental proposal by Vutha et al [18], as well as the theoretical advantages of the molecules, which is that they possess smaller values of $W_s/E_{eff}$ than those of the current leading candidates.

For numerical calculations, we employed an effective eEDM operator, $\hat{H}_{eEDM}^{eff}$ [19], given below, to reduce the computational cost.

$$\hat{H}_{eEDM}^{eff} = 2icd_e \sum_j^{N_e} \beta \gamma_5 \mathbf{p}_j^2 \quad (1)$$

Similarly, the operator for the S-PS interaction is given by [20,21]

$$\hat{H}_{S-PS} \equiv i\sqrt{2} G_F k_{s,A} Z_A \sum_j^{N_e} \beta \gamma_5 \rho_A(r_{Aj}) \quad (2)$$

We used Gaussian-type distribution function for $\rho_A$ similar to the reference [21]. In this paper, we discuss $W_s$ only for the heavy atoms (X = Sr, Cd, Ba, Yb, Hg, Ra and Cn). As mentioned earlier, we used the relativistic CCSD method to obtain the coupled cluster wavefunction, $|\Psi\rangle$ which is written using the reference wavefunction $|\Phi_0\rangle$ (the DF wavefunction in this case) as follows:

$$|\Psi\rangle = e^{\hat{T}} |\Phi_0\rangle \quad (3)$$

Here, $\hat{T}$ is the cluster operator. In the CCSD method, $\hat{T}$ is truncated as $\hat{T} = \hat{T}_1 + \hat{T}_2$, where $\hat{T}_1$ and $\hat{T}_2$ are single and double excitation operators, respectively. The details of the CCSD method are given elsewhere [22]. $E_{eff}$ and $W_s$ were calculated by using only the linear terms in the coupled cluster wavefunction as Eq. (4) shows, because the dominant contributions come from them.

$$\begin{aligned} E_{eff} &= -\frac{1}{d_e} \langle \Psi_{CCSD} | \hat{H}_{eEDM,N}^{eff} | \Psi_{CCSD} \rangle \\ &\approx \langle \Phi_0 | (1 + \hat{T}_1 + \hat{T}_2)^\dagger \hat{H}_{eEDM,N}^{eff} (1 + \hat{T}_1 + \hat{T}_2) | \Phi_0 \rangle_C \\ &+ \langle \Phi_0 | \hat{H}_{eEDM}^{eff} | \Phi_0 \rangle \end{aligned}$$
(4)

The subscript $N$ means that the operator, $\hat{H}_{eEDM,N}^{eff}$ is normal ordered, and $C$ means that only the connected terms are taken into account [23]. $W_s$ can be evaluated in a similar manner.

The DF and the CCSD computations were carried out in the modified UTChem [19,21,24] and the Dirac08 [25] codes, respectively. We used the multi-configurational Dirac-Fock method in the GRASP2K package to obtain the valence $s$ and $p$ orbitals for the heavier of the two atoms in each molecule, in the chosen set of molecules. We chose uncontracted, kinetically balanced [26] Gaussian Type Orbitals (GTOs). For the heavy atoms, we employed Dyall's triple zeta (TZ) basis sets [27] with polarization functions, while for F, we used Watanabe's basis set [28], with polarization functions from Sapporo basis set [29]. Cn is an exception, where we chose a Dyall's double zeta (DZ) basis set [30] without the polarization functions to avoid convergence problems. The details of the basis sets are shown in Supplemental Material [31]. We chose the following values of bond lengths (in Angstroms) for the molecules: SrF: 2.075 [32], CdF: 1.991 [33], BaF: 2.16 [32], YbF: 2.0161 [32] HgF: 2.00686 [34], RaF: 2.24 [35], and CnF: 2.070 [30]. In the calculation of $W_s$, we obtained the root-mean-squared nuclear charge radiuses, by using eq. (20) in [36]. We chose the following mass numbers for the calculations: Sr: 88, Cd: 114, Ba: 138, Yb: 174, Hg: 202, Ra: 223, and Cn: 285. In the CCSD calculations, we cut off the virtual spinors with orbital energy above 80 a.u..

Table I gives the results of our calculations for $E_{\text{eff}}$, $W_s$ and $W_s/E_{\text{eff}}$, both at the DF and the CCSD levels (we show only the absolute values for simplicity). The absolute values of both the properties increase with the atomic number of the heavier atom ($Z$), except in the case of CdF and HgF. The values for $E_{\text{eff}}$ for CdF and HgF are about twice as large as that for BaF and RaF respectively, even though Ba and Ra have a larger $Z$ values than Cd and Hg respectively.

The tendencies of $E_{\text{eff}}$ and $W_s$ calculated at the DF and CCSD method are same, in that the dominant contribution to the final result at the CCSD level comes from the DF term. Hence, we focus on the DF states and analyze the contribution from singly occupied molecular orbitals (SOMO). At the Kramers restricted DF level, only the SOMO contributes to the values of $E_{\text{eff}}$ and $W_s$. We show the results of Mulliken population (MP) analysis [37] for the SOMOs in Table II. MP indicates the contributions of each atomic orbital to SOMO. From this table, MPs of $s$ orbitals belonging to the heavier atom are by far the largest, followed by the contributions of the $p_{1/2}$ orbitals. The reason why the contributions of the $p_{1/2}$ orbitals are larger in group 12 molecules can be explained by the orbital interaction theory [38]. However, the larger $s$-$p$ mixings are not the main reason for the larger values of $E_{\text{eff}}$ and $W_s$ in group 12, as shown in the reference [39]. The $s$ and $p$ orbitals of the heavier atoms mainly contribute to the properties, as shown in Table III and IV and we do not discuss the other orbitals, because their contributions are negligible.

Table III and IV show the DF contributions to $E_{\text{eff}}$ and $W_s$, arising from the mixing of the heavier atom's $s_{1/2}^L$ (i.e. large component of $s_{1/2}$) and the $p_{1/2}^S$ (i.e. small component of $p_{1/2}$) basis sets, and the $p_{1/2}^L$ and $s_{1/2}^S$ basis sets. For example, the terms of the former, denoted by, '$s_{1/2}^L - p_{1/2}^S$' in $E_{\text{eff}}$ and $W_s$ are calculated, respectively, as follows

$$-4ci\sum_{k}^{N_s^L}\sum_{l}^{N_p^S} C_{s,k}^{*L} C_{p,l}^{S} \left\langle s_{1/2,k}^L \left| \mathbf{p}^2 \right| p_{1/2,l}^S \right\rangle \quad (5)$$

$$2\sqrt{2}G_{\text{F}}Zi\sum_{k}^{N_s^L}\sum_{l}^{N_p^S} C_{s,k}^{*L} C_{p,l}^{S} \left\langle s_{1/2,k}^L \left| \rho(r) \right| p_{1/2,l}^S \right\rangle \quad (6)$$

Here, $k$ and $l$ are the labels for basis set spinors of $s_{1/2}^L$ and $p_{1/2}^S$ respectively, and $C$ is a molecular orbital coefficient of the SOMO. $Z$ and $\rho(r)$ are the nuclear charge and nuclear charge density of the heavier atom, respectively.

Since the summation of the two terms ($s_{1/2}^L - p_{1/2}^S$ and $p_{1/2}^L - s_{1/2}^S$) is close to the DF value, and since the summation of the other contributions cancel in a way that they do not add up significantly, we can conclude that these two terms dominantly contribute to it. In the case of $E_{\text{eff}}$, there are large cancellations between the two terms, and the value that remains determines its total value. In contrast, in the case of $W_s$, the contribution of $s_{1/2}^L - p_{1/2}^S$ is always dominant, and this can be interpreted as being due to the form of the operator, $W_s$ that contains the nuclear charge density. Treating the nuclear charge density as a delta function at the nuclear region, we can see that only the $s_{1/2}^L - p_{1/2}^S$ term survives, since the large component of $s_{1/2}$ and the small component of $p_{1/2}$ both have overlap at the nucleus, unlike the $p_{1/2}^L - s_{1/2}^S$ term, where the latter does not overlap with the nucleus. Thus, in the case of $s_{1/2}^L - p_{1/2}^S$, the overlap between nuclear charge distribution and electronic wavefunction becomes much larger than the other combinations. This result is consistent with the discussion in a textbook [40], which mentions the behavior of the valence orbital in short-range is important.

As shown in Table I, $E_{\text{eff}}$ and $W_s$ do not always increase monotonically with $Z$. To explain this tendency, we analyze the valence atomic orbitals of the heavier atoms in the vicinity of the nucleus. In order to understand the distributions of the valence orbitals, we calculated the radial functions of the valence $ns$ and $np_{1/2}$ orbitals of the heavier atoms by using the GRASP2K package [17]. We observe that the order of the contraction of the wavefunction in descending order is Hg, Ra, Yb, Cd, Ba, and Sr, in the regions close to the nucleus, and this is of the same order of $E_{\text{eff}}$ and $W_s$ (The radial functions of large and small components are shown in Figures 1-4 in supplementary materials).

The contraction of the wavefunction in the nuclear region is also experimentally validated, by considering the hyperfine fields ($H_{\text{hyp}}$) [41]. $H_{\text{hyp}}$ is a measure of the density of the unpaired electron in the nuclear region, and is directly proportional to the experimentally measurable hyperfine coupling constant. The Cd and Hg atoms have larger hyperfine fields ($H_{\text{hyp}}$) than Ba and Ra atoms respectively, as shown in Table V. The tighter contractions of the valence orbitals of Cd and Hg atoms can be explained by the weak screening effect of $4d$ and $5d$ electrons, respectively.

Next, we discuss the correlation between the *atomic* wavefunction inside the nucleus and the *molecular* $E_{\text{eff}}$ and $W_s$. For the analysis of $E_{\text{eff}}$, we used the form of the eEDM Hamiltonian which includes $\mathbf{E}_{\text{int}}$ explicitly [43]. This is rewritten in the following way:

$$\sum_{j}^{N_{\text{elec}}} \beta_j \mathbf{\Sigma}_j \cdot \mathbf{E}_{\text{int}} = \sum_{j}^{N_{\text{elec}}} \left[ \mathbf{I}_4 + (\beta - \mathbf{I}_4) \right] \mathbf{\Sigma}_j \cdot \mathbf{E}_{\text{int}} \quad (7)$$

where $\mathbf{I}_4$ is a four-dimensional unit matrix. The expectation value of the first term on the RHS of Eq. (7) is zero [44], while the second term, which contains the coupling of small-small components, lead to a non-zero expectation value. We already showed in Table III that the heavier atom's $s_{1/2}$ and $p_{1/2}$ orbitals dominantly constitutes to the bulk of $E_{\text{eff}}$, at the DF level. And, since $E_{\text{eff}}$ is sensitive in the near-nuclear region, it is reasonable to look for how the property correlates with the orbitals in that region. Therefore, in Fig. 1, we plot $E_{\text{eff}}$ vs $Q_{ns}(r')Q_{np}(r')$, which is the product of the radial part of the heavier atom's valence $s_{1/2}^S$ and $p_{1/2}^S$, at the nuclear region. In this case, we take the value of $r' = 1.03 \times 10^{-7}$ (a. u.), as the closest point to the center of the nucleus calculated using GRASP2K. The radial functions were obtained using the GRASP2K program.

We observe from Fig. 1 that $E_{\text{eff}}$ correlates with the $s_{1/2}^S$ and $p_{1/2}^S$ orbitals in the nuclear region. In the case of $W_s$, we observe that the electronic wavefunction of $s_{1/2}^L$ and $p_{1/2}^S$ correlates to the expectation value, shown in Fig. 2. This correlation indicates that the tendency of *molecular* $E_{\text{eff}}$ and $W_s$ could possibly be explained by the *atomic* orbitals in the nuclear region. This line of reasoning is what links $E_{\text{eff}}$ with the weak screening effects in the nuclear region, enabling us to attribute the large values in group 12 monofluorides to the weak screening effects. This argument can be extended to $W_s$, too.

We also mention that the atomic enhancement factors $K$ for relevant paramagnetic cations (Group 12 and 2) show different trends, as compared to the molecular properties, $E_{\text{eff}}$ and $W_s$. The values of $K$ (41.2, 111.6, 315.0 and 812.5 for $Cd^+$, $Ba^+$, $Hg^+$ and $Ra^+$, respectively), obtained by using the DF method and the QZ basis sets, show that they increase with $Z$. This is due to the difference between the $s$-$p$ mixings in the atomic and molecular cases [39].

We can now explain the reason for the large values of $E_{eff}$ and $W_s$ in group 12 fluorides (CdF and HgF), using the idea of the weak screening effect of $(n-1)d$ electrons. Extending our reasoning to CnF, we find that the values of $E_{eff}$ and $W_s$ at the CCSD level are 662 GV/cm and 3360 kHz respectively, which are much larger than those of any other XF molecules. A detailed account of the higher relativistic effects in $d$-block elements are given in the references [45,46].

Although the values of $E_{eff}$ and $W_s$ in the group 12 monofluorides are larger than those of their group 2 counterparts, the ratio $W_s/E_{eff}$ monotonically increases with Z, as shown in Table I. This means that the weak screening effects do not seem to contribute to their ratios. The larger the difference between the ratios $W_s/E_{eff}$ for two given molecules, the cleaner the separation of $d_e$ and $k_s$, and therefore, one could preferably choose a lighter and a heavier system, in order to determine the individual values of $d_e$ and $k_s$. The present leading candidates, ThO, YbF and HfF$^+$, are heavy systems. We would have to focus on lighter systems.

Based on these findings, we propose SrF and CdF as new candidate molecules for experiment. Their effective electric fields and S-PS constants are not very large. However, they possess two major advantages. The first is in view of recent experimental proposal by Vutha *et al*, where they propose to embed the molecules in a solid matrix of inert gas atoms [18]. Their approach can result in statistical sensitivities that are far beyond the current best values, due to the large number of molecules that can be trapped, as well as the long precession times, for each molecule. The second advantage is a natural 'fine tuning'; these molecules have smaller values of $W_s/E_{eff}$ (see Table I) than ThO (2.84) [47], YbF (1.76, see Table I) and HfF$^+$ (1.75) [48], but $E_{eff}$ and $W_s$ are not small enough that SrF and CdF are no longer good eEDM candidates. For example, the $E_{eff}$ of CdF is larger than BaF, on which an eEDM experiment is currently under preparation [49], and the reason for its larger $E_{eff}$ is due to weak screening effects, as explained earlier. In addition, the smaller computational costs for SrF and CdF, as compared to heavier systems, are useful to us, since we can incorporate very accurately relativistic and correlation effects, as well as employ larger basis, and it becomes relatively much easier to get accurate results on $E_{eff}$ and $W_s$.

In summary, we find that group 12 molecules have larger values of $E_{eff}$ and $W_s$ than group 2 molecules. The reason for this is that the valence $s$ and $p$ orbitals of Cd and Hg are contracted more than their counterparts, Ba and Ra, respectively. These contractions are due to the weak screening effects of the $(n-1)d$ electrons in group 12 atoms. The ratio $W_s/E_{eff}$ increases with Z, and this behavior is different from those of $E_{eff}$ and $W_s$. Based on these results, we propose new candidate molecules, SrF and CdF. Both of them have a smaller $W_s/E_{eff}$ than ThO, YbF and HfF$^+$. The experiment for these molecules can be performed by embedding the molecules in a solid inert-gas matrix. These candidate molecules can shed light on the experiments for the determination of the value of $d_e$ and $k_s$.

**ACKNOWLEDGMENTS**

We would like to thank Dr. T. Aoki, for valuable discussions. We also wish to thank Dr. Amar Vutha, for his insightful points about the experimental aspects of CdF and SrF. This study was supported by the Core Research for Evolutional Science and Technology (CREST) program from the Japan Science and Technology (JST) Agency. This work was supported by JSPS KAKENHI Grant Number 17H03011, 17J02767 and 17H02881.

TABLE I. Summary of the calculated results, $E_{eff}$ (GV/cm), $W_s$ (kHz) and $W_s$(kHz)/$E_{eff}$(GV/cm) at the Dirac-Fock and CCSD levels.

|  | SrF | | CdF | | BaF | | YbF | | HgF | | RaF | |
|---|---|---|---|---|---|---|---|---|---|---|---|---|
| method | DF | CCSD | DF | CCSD | DF | CCSD | DF | CCSD | DF | CCSD | DF | CCSD |
| $E_{eff}$ | 1.5 | 2.1 | 9.1 | 10.9 | 4.8 | 6.6 | 18.2 | 22.3 | 105.3 | 114.9 | 42.2 | 54.5 |
| $W_s$ | 1.4 | 1.9 | 9.9 | 12.1 | 6.2 | 8.4 | 31.7 | 39.0 | 237.6 | 264.7 | 114.2 | 146.8 |
| $W_s/E_{eff}$ | 0.93 | 0.90 | 1.09 | 1.11 | 1.29 | 1.27 | 1.74 | 1.75 | 2.26 | 2.30 | 2.71 | 2.69 |

TABLE II. Mulliken Population of SOMO in XH molecules. Only the $s$ and $p_{1/2}$ components of the heavier atom are shown.

|  | SrF | | CdF | | BaF | | YbF | | HgF | | RaF | |
|---|---|---|---|---|---|---|---|---|---|---|---|---|
|  | large | small | large | small | large | small | Large | small | large | small | large | small |
| $s$ | 0.89 | $1\times10^{-5}$ | 0.79 | $3\times10^{-5}$ | 0.92 | $1\times10^{-5}$ | 0.85 | $2\times10^{-5}$ | 0.72 | $4\times10^{-5}$ | 0.95 | $2\times10^{-5}$ |
| $p_{1/2}$ | 0.14 | $2\times10^{-6}$ | 0.19 | $6\times10^{-6}$ | 0.11 | $2\times10^{-6}$ | 0.13 | $3\times10^{-6}$ | 0.17 | $9\times10^{-6}$ | 0.08 | $2\times10^{-6}$ |

Table III. Decomposed DF matrix elements of $E_{eff}$ (GV/cm) into the contribution of the large and small components of $s_{1/2}$ and $p_{1/2}$ spinors.

| Large-Small | SrF | CdF | BaF | YbF | HgF | RaF |
|---|---|---|---|---|---|---|
| $s_{1/2}^L - p_{1/2}^S$ | -18.6 | -71.6 | -26.2 | -61.1 | -266.4 | -83.7 |
| $p_{1/2}^L - s_{1/2}^S$ | 20.0 | 80.9 | 31.0 | 79.1 | 372.9 | 125.7 |
| Summation of above terms | 1.5 | 9.2 | 4.8 | 18.0 | 106.5 | 42.0 |
| Total value (DF) | 1.5 | 9.1 | 4.8 | 18.2 | 105.3 | 42.2 |

Table IV. Decomposed DF matrix elements of $W_s$ (kHz) into the contribution of the large and small components of $s_{1/2}$ and $p_{1/2}$ spinors.

| Small-Large | SrF | CdF | BaF | YbF | HgF | RaF |
|---|---|---|---|---|---|---|
| $s_{1/2}^L - p_{1/2}^S$ | 1.4 | 10.0 | 6.2 | 32.6 | 246.9 | 120.2 |
| $p_{1/2}^L - s_{1/2}^S$ | -0.0 | -0.1 | -0.1 | -0.9 | -9.4 | -5.8 |
| Summation of above terms | 1.4 | 9.9 | 6.1 | 31.7 | 237.5 | 114.4 |
| Total value (DF) | 1.4 | 9.9 | 6.2 | 31.7 | 237.6 | 114.2 |

Table V. Experimental $H_{hyp}$ (Tesla) of X atom cations.

|  | $Sr^+$ | $Cd^+$ | $Ba^+$ | $Yb^+$ | $Hg^+$ | $Ra^+$ |
|---|---|---|---|---|---|---|
| $H_{hyp}$ | 267[a] | 826[a] | 423[a] | 840[b] | 2626[a] | 1226[a] |

[a][41], [b][42]

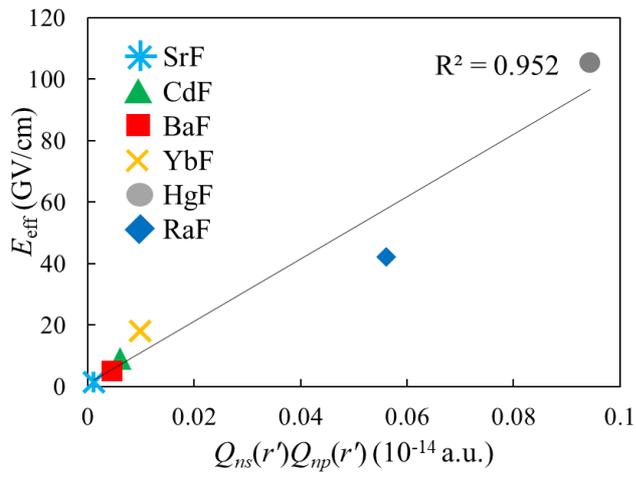

FIG. 1. Correlation between $Q_{ns}(r')Q_{np}(r')$ and $E_{\text{eff}}$.

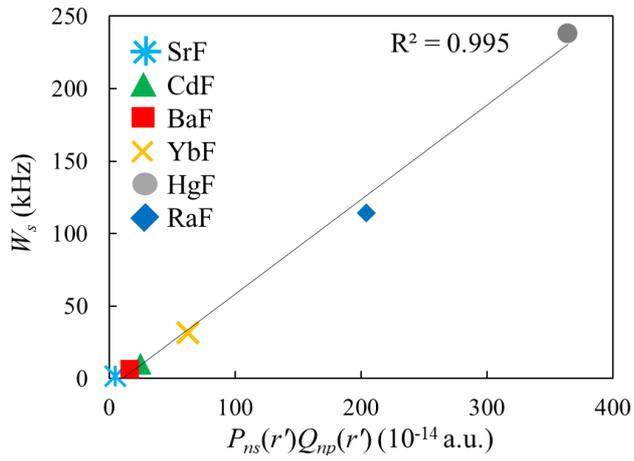

FIG. 2. Correlation between $P_{ns}(r')Q_{np}(r')$ and $W_s$.

# Supplemental Material

*ns* orbitals were evaluated from the ground states of the atoms. *np* orbitals were evaluated from the excited state of the atoms whose valence electron configurations are $ns^1np^1$. $P_{ns}(r)$ and $Q_{np}(r)$ are defined as follows.

$$\psi = \frac{1}{r}\begin{pmatrix} P^{\kappa}_{nl_L}(r)\,\chi^{\kappa}_m(\mathbf{r}/r) \\ iQ^{\kappa}_{nl_S}(r)\,\chi^{-\kappa}_m(\mathbf{r}/r) \end{pmatrix} \tag{S1},$$

where $\psi$ is an orbital of an atom and $\chi^{\kappa}_m(\mathbf{r}/r)$ is a spherical harmonics spinor.

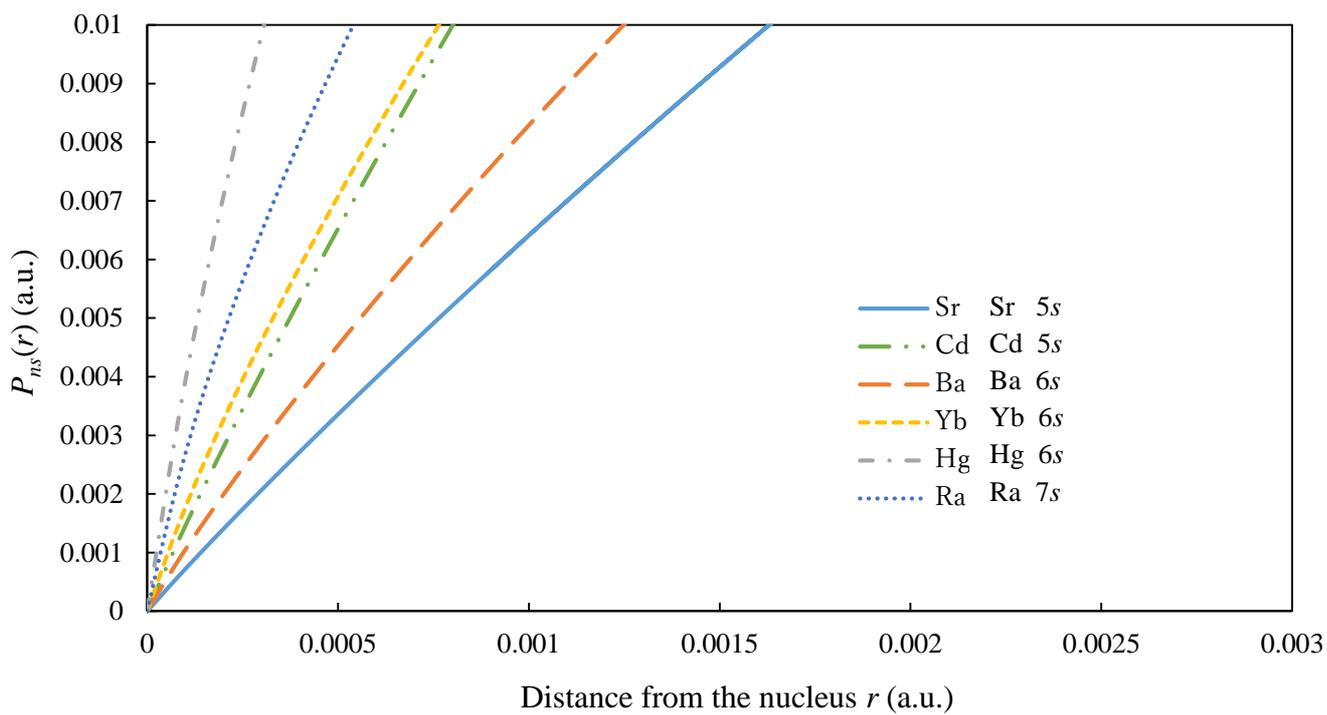

FIG. S1. Radial function of large component $P(r)$ of atomic valence $s$ orbital.

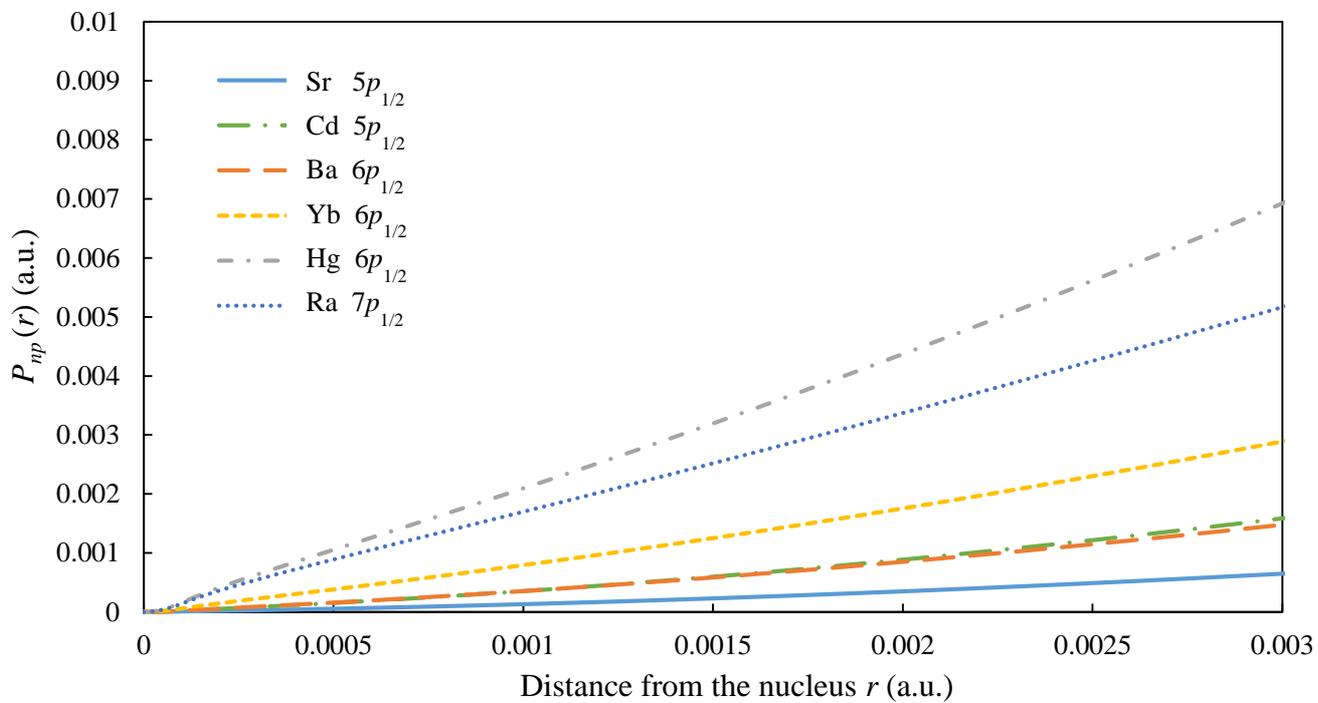

FIG. S2. Radial function of large component $P(r)$ of atomic valence $p_{1/2}$ orbital.

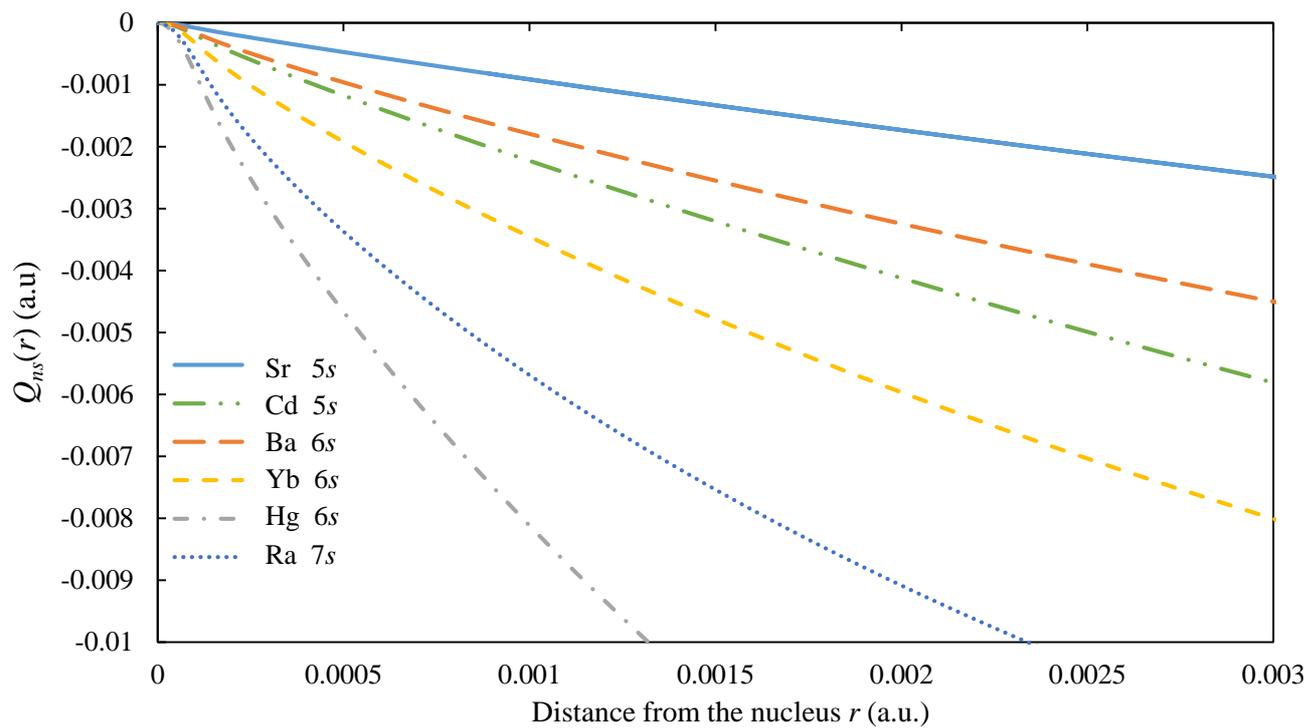

FIG. S3. Radial function of small component $Q(r)$ of atomic valence $s$ orbital.

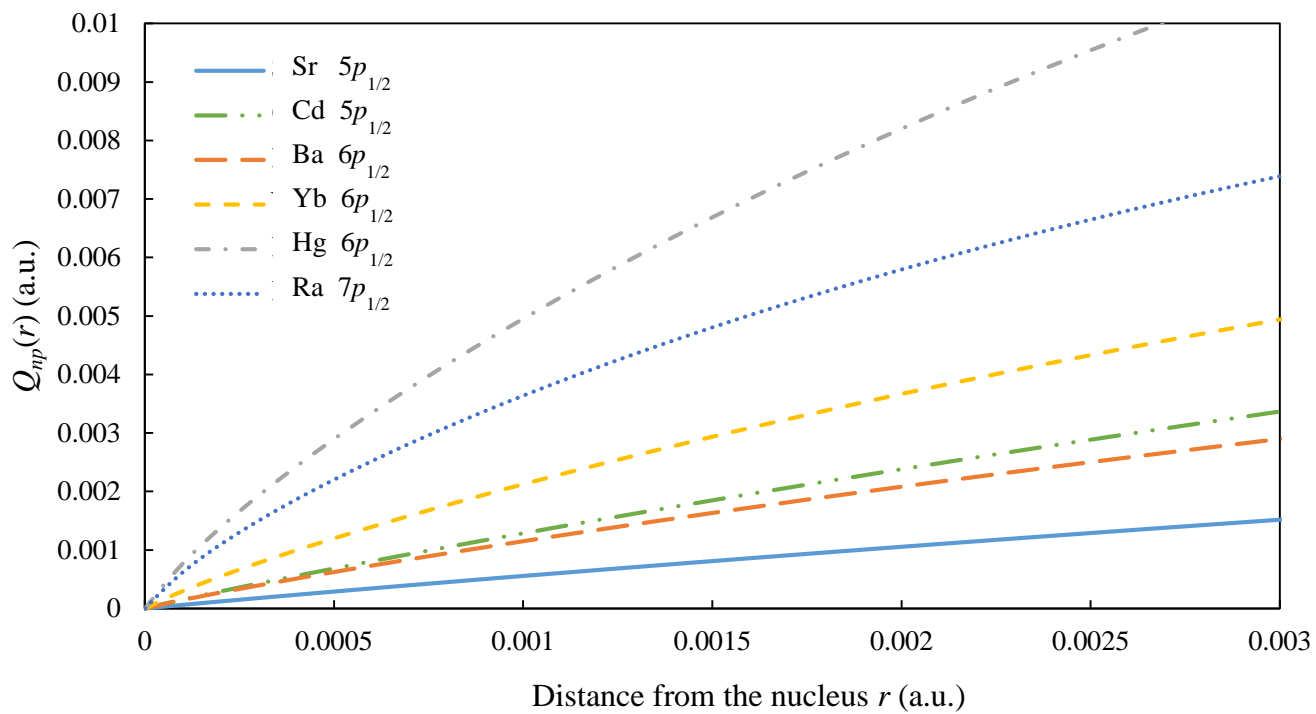

FIG. S4. Radial function of small component $Q(r)$ of atomic valence $p_{1/2}$ orbital.